\documentclass[a4paper]{article}

\usepackage{INTERSPEECH2021}

\usepackage{cite}
\usepackage{amsmath,amssymb,amsfonts}
\usepackage{algorithmic}
\usepackage{graphicx}
\usepackage{authblk}
\usepackage[T1]{fontenc}
\usepackage[utf8]{inputenc}
\usepackage{textcomp}
\usepackage{xcolor}
\usepackage{multicol}
\usepackage[frozencache,cachedir=.]{minted}
\usepackage{tcolorbox}
\usepackage{float}
\usepackage{hyperref}

\newtcolorbox{codebox}{
    breakable,
    colback=white!95!blue,
    colframe=black!75!blue,
    }

\title{Audiosockets: A Python socket package for Real-Time Audio Processing}
\name{Nicolas Shu$^\dagger$, David V. Anderson$^\dagger$}
\address{
  $^\dagger$Department of Electrical and Computer Engineering, Georgia Institute of Technology
  }
\email{nicolas.s.shu@gmail.com, anderson@gatech.edu}

\begin{document}

\renewcommand{\floatpagefraction}{.9}
\renewcommand{\topfraction}{.8}

\maketitle
\begin{abstract}
  
  There are many packages in Python which allow one to perform real-time processing on audio data. Unfortunately, due to the synchronous nature of the language, there lacks a framework which allows for distributed parallel processing of the data without requiring a large programming overhead and in which the data acquisition is not blocked by subsequent processing operations. This work improves on packages used for audio data collection with a light-weight backend and a simple interface that allows for distributed processing through a socket-based structure. This is intended for real-time audio machine learning and data processing in Python with a quick deployment of multiple parallel operations on the same data, allowing users to spend less time debugging and more time developing.
  
\end{abstract}
\noindent\textbf{Index Terms}: python, audio, speech, socket programming, real-time processing

\section{Introduction}

Machine learning and analysis of audio signals is a rapidly growing field.   Although most of the machine learning research for audio processing is done offline, practical systems often require on-line, real-time operation.  There are many packages available in Python for audio loading \cite{pyaudio, scipy, librosa, soundfile, sounddevice}, and a few which allow one to record audio in real-time \cite{pyaudio, sounddevice}. 
However, often it is the case in a synchronous language like Python, where a process that is continuously recording audio may only record, and cannot asynchronously spend its compute time processing the audio through an algorithm (e.g. a neural network). There exist solutions such as using Robot Operating System (ROS) \cite{ros} to perform message passing between parallel processes; but,  although easy to use and very useful for multiple tasks, ROS is a relatively large software which contains multiple features that may be too heavyweight for a simpler task which only requires message passing.
Currently, there lacks an off-the-shelf software package for Python which is simple to use and quickly deployable for real-time processing for Python-related audio projects. This work introduces audiosockets, a socket-based package which uses the Sounddevice package \cite{sounddevice} to record audio, deploys a local server and allows for multiple clients to behave as processors for the incoming audio. This package handles all of the socket programming in the backend and is designed to minimize the user's efforts on  designing the message passing, and maximize their time developing new algorithms. 

\section{Related Work}
While there are packages specifically centered around to do audio processing such as PySox and Librosa, there are a few options which one may take to perform real-time inference over audio data. PyAudio is an example of a package which may be used for real-time inference. It allows a user to open audio streams and contains a callback mode, which allows a user to place the data, which is stored as byte strings, in a queue without blocking the script, making it a very powerful tool for a programmer. The one minor downside of PyAudio is that it requires a good understanding of how microphones operate, and it requires a significant amount of  programming and attention to detail before getting the code to function properly. Furthermore, the PyAudio package by itself does not allow one to distribute the post-processing work across different threads. 

PyJack is another package which interfaces with the JACK (JACK Audio Connection Kit) API, however, the code was originally written for Python 2.7, and, at the time of this writing, the latest release was in March 2011, leaving many of the connections deprecated. PyGame is a solution that has a similar interface as PyAudio, where it requires one to create an object which will use a callback function to append chunks of data in bytes format to a list, but it also suffers from the lack of work distribution across different threads for post-processing the data. Sounddevice is another feasible solution as it is also capable of using callback functions, allowing one to store the data in a queue, but it also already converts the output data as numerical values, ready for other use cases.

\section{Method}

\begin{figure*}
  \includegraphics[width=\textwidth]{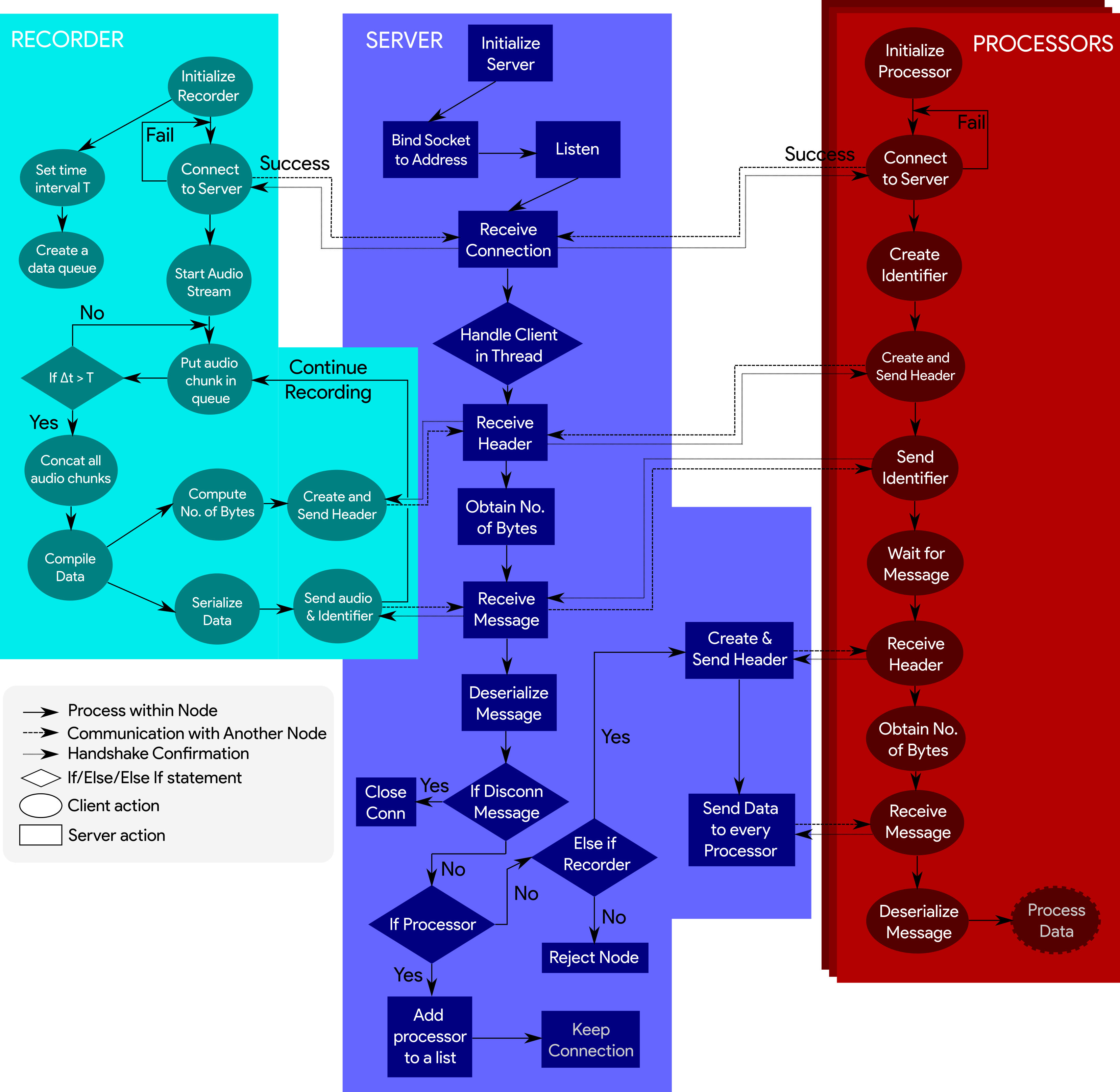}
  \caption{Mechanism for the audiosockets package. The different colored superblocks represent the different parallel processes which are operating and communicating with each other. The superblocks containing circular processes (i.e. Recorder and Processors) are the clients, and the superblock with rectangular processes represents the local server which is to be deployed in a system.}
  \label{fig::mechanism}
\end{figure*}

This package uses socket programming as its backend communication protocol and the package Sounddevice \cite{sounddevice} to obtain streams of audio from a microphone as well as NumPy \cite{numpy}. Therefore, the only packages needed to be installed are sounddevice and NumPy, as well as PortAudio for your operating system. The architecture for this package is shown in Figure \ref{fig::mechanism}.

This package deploys a local server on one's system, which will bind a socket to an IP address and a pre-defined port, and will start to listen for new connections from clients. Given that a local area network exists, this package is capable of sending data over a network. Clients will have two different types: ``recorder'' and ``processor'', and the server will know how to handle each of those two types of clients. The two types of clients are seen as equal to the server, and through Python's threading module, there is a thread which waits for clients to connect to it while another thread starts to connect a new incoming connection.

For the case of a recorder, once initialized, the recording node will create a socket and will attempt to connect to the pre-defined address of the server. The node will attempt to connect to the server at 1Hz, and once it does connect, it will create a queue object and create a stream object which continuously dumps the audio data from the microphone in the queue. Periodically, it will start to send the audio data to the server. To do so, it will perform two major steps. The first step is to serialize the data and to determine the number of bytes required to send the message to the server. Once the number of bytes has been determined, the node will create a header message of size of 64 bytes, and will send a message header to the server specifying the number of bytes for the full audio data that the server is expected to receive. 

Once the server receives a new connection from the Recorder client, and it will wait for a new messages to be received. It will expect two messages. First, a message of size 64 bytes which specifies the number of bytes of the actual message to be received, while the second will be the actual message. Once the server receives the header, it will send back a message to the client stating that it has successfully received the header message, and it's ready to receive the data. The recorder will then send a message containing the all the data including: the audio data, the sampling frequency, the current time that the message was sent, and a type identifier for the server so that the server knows how to handle the given client. The server will then receive the data message, and send back a confirmation back to the client stating that it has received the data message. Since the stream of audio from Sounddevice works in parallel to the script, it will continue to gather more data, and then the recorder will repeat the recording steps until it is time to send a new message to the server. 

Unfortunately, in socket programming, just because a connection expects to receive a specific sized message, let us say 2048 bytes, it does not mean that it will receive the full message at once. Therefore, as the server receives the stream of data, it will continue to check for the rest of the fragments of the message until it receives the entire message, and only then will it send the confirmation back to the client that it has received the data message. At this point, the server first needs to deserialize the message onto an actual object. Once the object has been interpreted, the server will check the type of client that just sent the message. If the client was a recorder, it will follow a recording protocol. If the client was a processor, then it will follow a processor protocol. But if it was none of the above, it will send an error message back to the client and dump the data. 

If the server receives a data message from a processor, the first thing it does is to register the processor to the server by name on a pointer, and it holds that connection opened. If the server receives a data message from a recorder, then it sends the data to all of the registered processor connections. 

The processor client behaves in a very simple manner. Once it is initialized, as described before as a characteristic of it being a client, it attempts to connect to the server. Once a connection is established, the processor node sends a header message specifying the size of the data message that it will send the server, and once the server has confirmed that it has received the message, the processor node will send a data message containing a name identifier and a type identifier. As described before, the processor node will be registered to the server as a processor, and then it will block itself waiting for a header message from the server. Once it receives a header message from the server, it will send a confirmation back to the server that it has received the header message, and then will wait to receive a data message, which will, in turn, be deserialized and interpreted by the object. The object will then have a method that will process the data in whichever way the user desires. 

To deal with disconnections, all of the clients have methods that will send disconnecting data messages to the server. If the server receives a disconnecting message from a client, it will delete the processor from the registered processors (if the client is a processor), the server will send a confirmation message back to the client that it will disconnect the client, and then proceed to close the connection between the server and the client. 

\begin{figure}[h]
  \includegraphics[width=\linewidth]{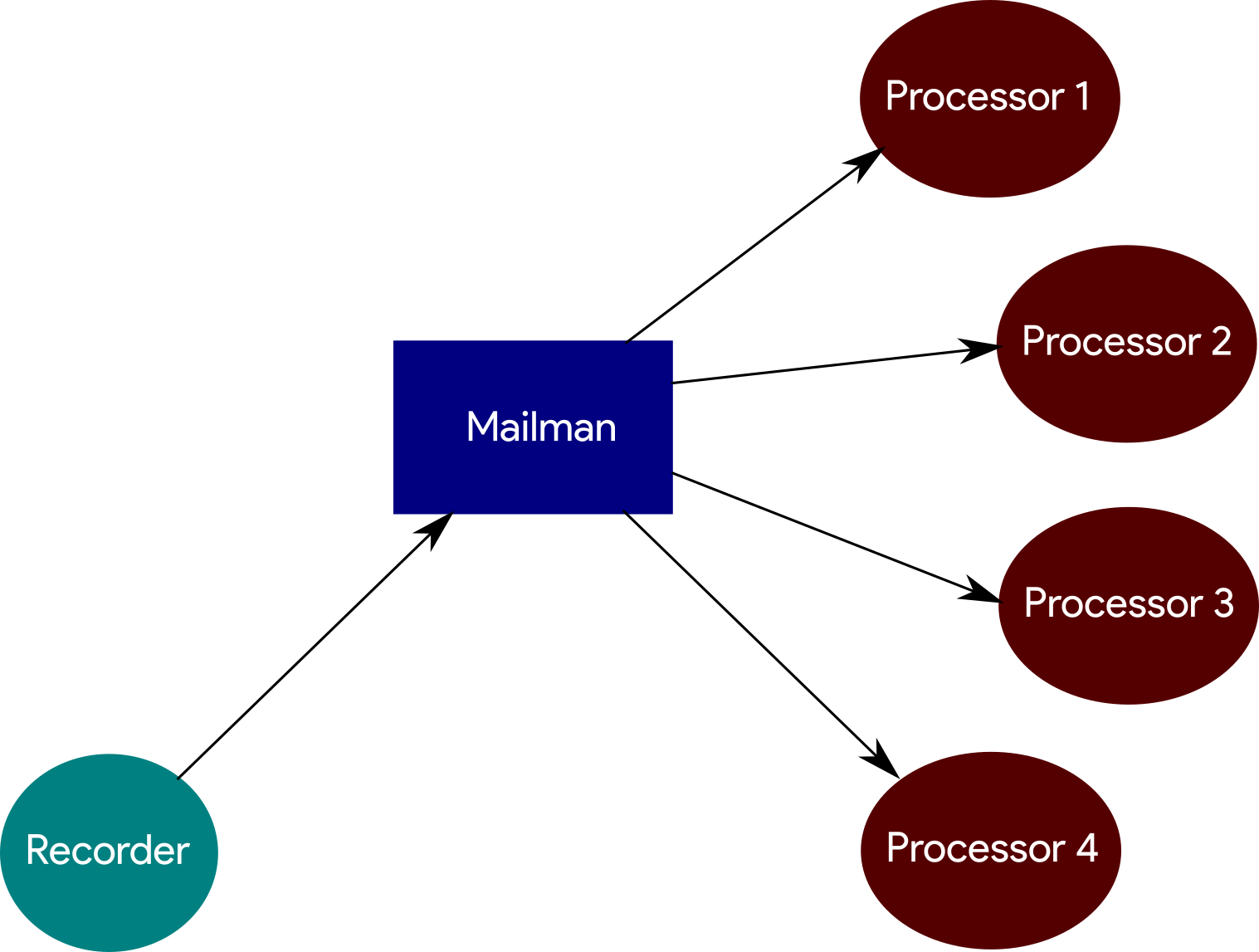}
  \caption{A simplistic overview diagram of the capabilities of the package}
  \label{fig::multiple_processors}
\end{figure}

The versatility of this method allows for one to connect to multiple processors at once and to disconnect any of the clients (processors or recorder) at any point in time, and reconnect it again, as shown in Figure \ref{fig::multiple_processors}

\section{Usage}
Here, a baseline is described on how to use this package. 

\subsection{Network Setup}
In order to setup the network, one needs to create a JSON file which contains a few parameters
\begin{itemize}
    \item \texttt{``SERVER''}: The string IP address of the machine where server is deployed to
    \item \texttt{``PORT''}: The port number in the local computer where the server will be deployed and the port where the clients will try to bind to. 
    \item \texttt{``HEADER''}: The number of bytes defined for size of the header message
    \item \texttt{``FORMAT''}: The type of which text data will be encoded or decoded
    \item \texttt{``DISCONNECT\_MSG''}: The disconnecting message which the clients will send to the server and the message that the server will be expecting to disconnect the client
    \item \texttt{``logging\_format''}: The format for the logging module to report the logs
    \item \texttt{``logging\_level''}: The level of logging to be displayed
\end{itemize}

Thus an example is as shown below
\begin{minted}[fontsize=\footnotesize]{json}
{
  "SERVER": "172.16.12.10",
  "PORT": 5050,
  "HEADER": 64,
  "FORMAT": "utf-8",
  "DISCONNECT_MSG": "DISCONNECT",
  "logging_format": 
    "%(asctime)s-%(message)s", 
  "logging_level": "info"
}
\end{minted}

\subsection{Server Deployment}
In order to deploy a server, one needs to import the \mintinline{python}{MailmanSocket}, pass the path to the JSON file, and run the \mintinline{python}{start()} method. An example is shown below:

\begin{minted}[fontsize=\small]{python}
from audiosockets import MailmanSocket
if __name__ == "__main__":
    mailman = MailmanSocket(
        "server_info.json")
    mailman.start()
\end{minted}

\subsection{Recorder Deployment}
In order to deploy the recorder, one  needs to import the \mintinline{python}{RecorderSocket}, pass the path to the JSON file specified above, and run the \mintinline{python}{start()} method. An example is shown below

\begin{minted}[fontsize=\small]{python} 
from audiosockets import RecorderSocket
if __name__ == "__main__":
    recorder = RecorderSocket(
        "server_info.json")
    recorder.start()
\end{minted}

\subsection{Processors Deployment}
The processor is slightly more involved, but that is up to the user with respect to how complex does he or she wishes to process the data. First, one needs to import the class \mintinline{python}{BaseProcessorSocket}, and inherit from it. One must pass the arguments to the parent's \mintinline{python}{__init__()} method, and then one must overwrite the method \mintinline{python}{process_data(data)} method, which takes in a single argument, with anything the user wishes to do with the data. The example below shows the usage of a LogMelSpectrogram class that computes the log Mel spectrogram of a signal given a sampling frequency. Finally, one must run the method \mintinline{python}{start()} so that it binds to the server and becomes part of the network. \cite{soundfile}

\begin{minted}[fontsize=\small]{python}
from audiosockets import BaseProcessorSocket
from audiosockets.utils import \
    LogMelSpectrogram as LMSProcessor
class LMSProcessor(BaseProcessorSocket):
    def __init__(self,*args, **kwargs):
        super().__init__(*args, **kwargs)
    def process_data(self,data):
        fs = data["fs"]
        audio = data["data"]
        lms = LogMelSpectrogram(fs)(audio)
        print(lms.shape)
if __name__ == "__main__":
    processor = LMSProcessor(
        "VAD", 
        "server_info.json")
    processor.start()
\end{minted}


\section{Conclusion}

Before this work, there lacked an off-the-shelf open-source framework that would allow one to quickly deploy real-time audio processing in Python. For some solutions, a lot of understanding with regards to lower-level analog-to-digital processing was required. Other solutions would hit a roadblock due to the synchronous behavior of Python, where, for a real-time application, if an algorithm takes longer to process the data than the audio collection itself, it then becomes the bottleneck to the pipeline, and data may potentially be lost. Furthermore, they would all require additional work to get a distributed computing framework where different nodes would contain different algorithms, and would not bottleneck the node collecting the data. This work fills all of the gaps previously mentioned. It allows one to quickly deploy a real-time audio processing solution, where a node is responsible for only collecting data and sending it to a local server, and the local server's role is to simply receive any data the first node first sent it and sends it to other nodes which each may contain an algorithm which likely does not require to be re-initialized. As an example, neural networks require, relatively speaking, a long time to be initialized and prepared for inference. Depending on the capabilities of the machine, this framework allows the user to deploy several different neural networks on their GPU, each of which may be specialized for a different task, and the local server sends the data to each of them. At no point is there a loss of data, and with a minimal amount of work, a fully functional real-time processor can be deployed. 

\section{Acknowledgements}

We like to thank Mouhyemen Khan for the fruitful discussions over ideas which could be used in order to implement and achieve the goal of this work.

\bibliographystyle{IEEEtran}

\bibliography{mybib}

\begin{thebibliography}{1}
\providecommand{\url}[1]{#1}
\csname url@samestyle\endcsname
\providecommand{\newblock}{\relax}
\providecommand{\bibinfo}[2]{#2}
\providecommand{\BIBentrySTDinterwordspacing}{\spaceskip=0pt\relax}
\providecommand{\BIBentryALTinterwordstretchfactor}{4}
\providecommand{\BIBentryALTinterwordspacing}{\spaceskip=\fontdimen2\font plus
\BIBentryALTinterwordstretchfactor\fontdimen3\font minus \fontdimen4\font\relax}
\providecommand{\BIBforeignlanguage}[2]{{%
\expandafter\ifx\csname l@#1\endcsname\relax
\typeout{** WARNING: IEEEtran.bst: No hyphenation pattern has been}%
\typeout{** loaded for the language `#1'. Using the pattern for}%
\typeout{** the default language instead.}%
\else
\language=\csname l@#1\endcsname
\fi
#2}}
\providecommand{\BIBdecl}{\relax}
\BIBdecl

\bibitem{pyaudio}
\BIBentryALTinterwordspacing
``Pyaudio: Cross-platform audio i/o for python, with portaudio.'' [Online]. Available: \url{https://people.csail.mit.edu/hubert/pyaudio/}
\BIBentrySTDinterwordspacing

\bibitem{scipy}
P.~Virtanen, R.~Gommers, T.~E. Oliphant, M.~Haberland, T.~Reddy, D.~Cournapeau, E.~Burovski, P.~Peterson, W.~Weckesser, J.~Bright, S.~J. {van der Walt}, M.~Brett, J.~Wilson, K.~J. Millman, N.~Mayorov, A.~R.~J. Nelson, E.~Jones, R.~Kern, E.~Larson, C.~J. Carey, {\.I}.~Polat, Y.~Feng, E.~W. Moore, J.~{VanderPlas}, D.~Laxalde, J.~Perktold, R.~Cimrman, I.~Henriksen, E.~A. Quintero, C.~R. Harris, A.~M. Archibald, A.~H. Ribeiro, F.~Pedregosa, P.~{van Mulbregt}, and {SciPy 1.0 Contributors}, ``{{SciPy} 1.0: Fundamental Algorithms for Scientific Computing in Python},'' \emph{Nature Methods}, vol.~17, pp. 261--272, 2020.

\bibitem{librosa}
B.~McFee, C.~Raffel, D.~Liang, D.~P. Ellis, M.~McVicar, E.~Battenberg, and O.~Nieto, ``librosa: Audio and music signal analysis in python,'' in \emph{Proceedings of the 14th python in science conference}, vol.~8, 2015.

\bibitem{soundfile}
{Bastian Bechtold}, ``Pysoundfile.''

\bibitem{sounddevice}
\BIBentryALTinterwordspacing
M.~Geier \emph{et~al.}, ``Sounddevice,'' 2020. [Online]. Available: \url{https://python-sounddevice.readthedocs.io/en/0.3.15/index.html}
\BIBentrySTDinterwordspacing

\bibitem{ros}
\BIBentryALTinterwordspacing
{Stanford Artificial Intelligence Laboratory et al.}, ``Robotic operating system.'' [Online]. Available: \url{https://www.ros.org}
\BIBentrySTDinterwordspacing

\bibitem{numpy}
\BIBentryALTinterwordspacing
C.~R. Harris, K.~J. Millman, S.~J. van~der Walt, R.~Gommers, P.~Virtanen, D.~Cournapeau, E.~Wieser, J.~Taylor, S.~Berg, N.~J. Smith, R.~Kern, M.~Picus, S.~Hoyer, M.~H. van Kerkwijk, M.~Brett, A.~Haldane, J.~F. del R{'{\i}}o, M.~Wiebe, P.~Peterson, P.~G{'{e}}rard-Marchant, K.~Sheppard, T.~Reddy, W.~Weckesser, H.~Abbasi, C.~Gohlke, and T.~E. Oliphant, ``Array programming with {NumPy},'' \emph{Nature}, vol. 585, no. 7825, pp. 357--362, Sep. 2020. [Online]. Available: \url{https://doi.org/10.1038/s41586-020-2649-2}
\BIBentrySTDinterwordspacing

\end{thebibliography}


\end{document}